Review article

Variations on a theme: changes to electrophoretic separations that can make a difference.

Thierry Rabilloud[1,2,3]

1: CNRS UMR5092, Biochemistry and Biophysics of Integrated Systems, CEA Grenoble, iRTSV/BSBBSI, 17 rue des martyrs, F-38054 GRENOBLE CEDEX 9

2 : CEA-DSV/iRTSV/LBBSI, Biophysique et Biochimie des Systèmes Intégrés, CEA-Grenoble, 17 rue des martyrs, F-38054 GRENOBLE CEDEX 9, France

3: Université Joseph Fourier, UMR CNRS-CEA-UJF 5092, CEA-Grenoble, 17 rue des martyrs, F-38054 GRENOBLE CEDEX 9

Correspondence to

Thierry Rabilloud, iRTSV/BBSI
CEA-Grenoble, 17 rue des martyrs,
F-38054 GRENOBLE CEDEX 9
Tel (33)-4-38-78-32-12
Fax (33)-4-38-78-44-99
e-mail: Thierry.Rabilloud@ cea.fr

Abstract

Electrophoretic separations of proteins are widely used in proteomic analyses, and rely heavily on SDS electrophoresis. This mode of separation is almost exclusively used when a single dimension separation is performed, and generally represents the second dimension of two-dimensional separations.
Electrophoretic separations for proteomics use robust, well-established protocols. However, many variations in almost all possible parameters have been described in the literature over the years, and they may bring a decisive advantage when the limits of the classical protocols are reached.
The purpose of this article is to review the most important of these variations, so that the readers can be aware of how they can improve or tune protein separations according to their needs.
The chemical variations reviewed in this paper encompass gel structure, buffer systems and detergents for SDS electrophoresis, two-dimensional electrophoresis based on isoelectric focusing and two-dimensional electrophoresis based on cationic zone electrophoresis.

1. Introduction

Strictly speaking, two-dimensional electrophoresis is the combination of any type of electrophoreses. However, in the vast majority of cases, the second dimension is fixed to SDS-PAGE. Consequently, in order to maximize resolution, the first dimension shall be as remote as possible, in its separation principles, from SDS-PAGE. Although this leaves quite a wide space for non-classical 2D gels, as reviewed in [1], the first separation of 2D gels generally falls into three main categories, namely: i) denaturing zone electrophoresis [2, 3] ,ii) native zone electrophoresis, as pioneered by Blue-Native PAGE [4] and in most cases iii) denaturing isoelectric focusing, as first described in [5, 6].

For all of these three basic setups, many changes and adaptations have been devised over the years to improve the performances of the two-dimensional electrophoretic separation and /or to adapt them to the experimental needs. The word performance can encompass resolution, scope of analytes, analysis window in terms of protein parameters, loading capacity, and any type of combination of these parameters.

Many of these optimized variants have been described in the literature, and are quite often forgotten, although many can still be very useful. The purpose of this review is therefore to redescribe some of these variants and bring them back to light, hoping that it will be useful to researchers using the various types of 2D gels for their research. However, some areas are deliberately left out of the scope of this review. One of these is the native/denaturing setup, which has been recently reviewed elsewhere [7]. The other is the pH gradient engineering in the case of IEF-based separations, as this too has been reviewed thoroughly [8]. Last but not least, the field of protein detection after electrophoresis will be left out of the scope of this review, and readers interested in this aspect should refer to a recent review [9].

Thus, this review will be focused on two types of SDS-PAGE-based 2D separations using respectively denaturing zone electrophoresis and denaturing IEF as the first separation.

2. Variations in SDS-PAGE

Technically speaking, a SDS-PAGE separation is made in a gel containing a buffer and a detergent. Not surprisingly, numerous variations have been attempted for all three parameters, and will therefore be described below

2.1. Variations in the gel structure

Compared to the classical acrylamide-bisacrylamide gel, which makes the vast bulk of SDS gels run nowadays, some variants have been devised to improve the characteristics of the gel itself. Polyacrylamide gels are relatively fragile, and this can be felt especially when large-size gels are used [10]. This fragility has indeed hampered the wide use of very large gels, despite their wonderful resolution, as resolution in 2D gels goes with the surface of the gel, i.e. the square of the linear dimensions. Thus, doubling the size of each dimension increases the resolution by a factor of 4.

To increase the mechanical resistance of the gels, two approaches have been described. In the first one [11], a linear polymer (linear, uncrosslinked polyacrylamide) is added to the monomer solution before polymerization, and the ready to use mixed solution was sold under the trade name Duracryl. Considerably enhanced mechanical properties were described. However, with the advent of the mass spectrometry-based protein identification techniques, the presence of a polymer in the gel considerably degraded the performances in terms of protein identification. This led to very limited use of this matrix nowadays.

The second way on increasing the mechanical strength of the gel is to change the monomers themselves. One successful attempt in this direction was made by changing the crosslinker of the gels, and namely to replace methylene-bisacrylamide by bis-acryloyl piperazine. Bis-acryloyl piperazine was first described in conjunction with another monomer, acryloyl morpholine [12], but its successful use in 2D gels was to be described some years later, and in conjunction with regular acrylamide as the main monomer [13]. In addition to greater mechanical properties, bis-acryloyl piperazine was also shown to provide slightly larger pores in the gel and improved performance in some silver staining techniques [13].

This story shows how difficult it is to end up with monomers that are really useful for protein separations in SDS PAGE. In fact, the higher the molecular mass of the monomer, the better in terms of flexibility, as at constant monomer mass percentage a gel with wider pores will be obtained, while at constant pore size more monomer will be added, resulting in a stronger gel. However, the story of acryloyl morpholine shows that the game is not that simple, as heavy monomers are generally more hydrophobic and either lead to phase separation during polymerization, with an acrylic plastic separated from the aqueous phase, or to gels that have strong hydrophobic properties and therefore perform very poorly in protein electrophoresis.

The solution to this problem would be to use heavy but hydrophilic monomers, such as acryloylaminosugars [14] or trisacryl [15]. However, the use of these monomers is not without problems, and points to a second important feature of polyacrylamide gels, which is their chemical resistance. After polymerization, polyacrylamide can be seen as a polyethylene backbone with numerous pending amido groups conferring to the polymer its hydrophilicity and thus its ability to form hydrogels (see figure 1). The polyethylene backbone is chemically very resistant, but the amido groups are sensitive to hydrolysis, especially under basic conditions. This leads to conversion of polyacrylamide into polyacrylate, with deeply altered properties, such as greater swelling in water, strong electroendosmosis and ion retention properties. These properties are unwanted for electrophoretic applications, and base-resistant monomers should be used if possible. Acrylamide is reasonably stable at the pH present during electrophoresis (see next section on buffers), but the highly hydroxylated monomers such as acryloylaminosugars and trisacryl are much more prone to base-catalyzed hydrolysis [16], and this precludes their use in SDS PAGE under standard conditions. Some monomers, however, have been developed specifically to combine hydrophilicity with resistance to hydrolysis, such as acrylamidoethoxyethanol (AAEE) [17] and acrylamidopropanol (AAP) [18]. The use of these monomers has remained confidential up to now, but a recent paper focused on membrane proteins takes advantage of the base resistance of AAP gels to use them in conjunction with in gel proteolysis at high pH with proteinase K [19].

Another feature of interest in gels used for proteomics applications is the control of the pore size. Ideally, the gel should have a small pore size during electrophoresis to perform an efficient separation, and then a larger pore size to facilitate protease penetration and peptide elution. This means to be able to cleave the crosslinker, ideally only part of the bonds to keep a solid gel structure and not to pollute the extracts with polymers. Regular acrylamide-bisacrylamide gels are quite resistant and require concentrated hydrogen peroxide to be cleaved [20]. These conditions would be of course completely incompatible with any decent peptide analysis. Thus, it is required to change the crosslinker for manipulating the pore size, and quite different cleavable crosslinkers have been developed over the years (reviewed in [21]). However, these crosslinkers are not always easy to use. Some (e.g. the allyl derivatives) polymerize very poorly [21], Others such as ethylene diacrylate and bisacrylyl cystamine, cleave under the conditions used for SDS PAGE, i.e. during the migration, which is of course to be avoided. Thus, the only successful reports in proteomic-type applications up to now use either DHEBA with subsequent base-catalyzed cleavage [22], or a new acid-cleavable crosslinker [23]. In the case of DHEBA, however, it must be stressed that this report used immunoblotting as the protein probing technique. As cleavage of DHEBA under basic conditions will liberate glyoxal (see figure 1), it can be expected that this technique will not be very useful for mass spectrometry-based proteomics setups.

This drives to the last point in the gel structure that may interfere with the subsequent proteomic analysis, i.e. the chemical reactivity of the gel itself on the proteins. A perfect polyacryamide gel would be completely unreactive toward proteins, but real-life polyacrylamide gels are not perfect. First of all, the conversion efficiency of the monomer into the polymer is far from perfect under standard conditions [24, 25]. Moreover, a 90% conversion efficiency in a 7.5% gel still means 0.1M of free acrylamide at the end of the polymerization process. It is therefore not surprising that, under the basic conditions prevailing in SDS PAGE, such a concentration of free acrylamide can lead to addition on the most nucleophilic amino acids, e.g. cysteine [26]. In fact, the extent of aminoacid alkylation observed in proteomics is relatively low when compared to this concentration of acrylamide, and this may be due to the use of a glycine-containing buffer in SDS PAGE, glycine acting as a scavenger for unpolymerized acrylamide [27].

Another problem in the use of classical polyacrylamide gels for proteomics lies in the chemicals used for inducing polymerization. What is commonly called "catalysts" (i.e. TEMED and persulfate) are not catalysts, because they are not recovered unchanged at the end of the process. Their proper naming is initiators, as they initiate the polymerization process. In fact, TEMED-derived species and persulfate-derived species can be found at both ends of the polymer chains, as they will graft covalently at the start of the chain to initiate polymerization (see figure 1), but can also play a role in the termination reactions. Thus, many chemicals used to initiate polymerization, although ionic, will not be migrated away during the electrophoretic process, as they are covalently bound to the gel itself.
This has been used purposely to graft thiosulfate in the gel in order to decrease the background during silver staining [28], but the key point is that standard TEMED-persulfate systems lead to gels that are oxidizing. This has been very clearly demonstrated for IPG gels [29], but it is highly likely that the same holds true (although probably to a lesser extent) in SDS gels [30], and may explain why methionine sulfoxides are so often observed in any type of gel-based proteomics.
Although not widely used, to say the least, there are however polymerization systems that alleviate most of these problems. These systems are based on photopolymerization. The first photopolymerization systems, based on riboflavin [30], were poorly efficient and rather unreliable. However, it could be demonstrated that photopolymerization did not induce any oxidizing power in the resulting gels [31].

However, improved photopolymerization systems based on methylene blue were also introduced [32] resulting in excellent monomer conversion in almost any condition of pH [33] and solvents [34]. It was shown, however, that this system is not fully compatible with SDS electrophoresis, due to the cationic nature of methylene blue [35]. The solution found was to replace methylene blue by riboflavin phosphate, but also keeping the co-initiators toluene sulfinate and diphenyliodonium chloride. This resulted in efficient and reliable polymerization [35], with decreased protein modification (as assessed by Edman sequencing). However, its is clear that such photopolymerization systems, replacing methylene blue with anionic dyes, result in less artefacts in the subsequent proteomic analyses [36].

Although these photopolymerization systems are cumbersome to use when large number of gels are to be run, as in 2D gels-based proteomics, they should bring some improvements in other proteomics setups using SDS-PAGE, such as the widely used GeLC system [37].

2.2. Variations in the buffer systems

Due to their high resolution, only discontinuous systems are used. In most cases, the original Tris-HCl-glycine system is used [38], although many systems can be devised by application of the discontinuous electrophoresis theory [39, 40]. A simple point that must be kept as a reminder for anionic systems, as those used for SDS-PAGE, is that for any given system, the mobility of the ions is pH-dependent. The higher the pH, the faster the moving boundary, and thus the more the system is able to resolve low molecular mass proteins, at the expense of the resolution space allocated to high molecular mass proteins. Conversely, the lower the pH, the more space is allocated to resolve high molecular mass proteins, with the low molecular mass ones comigrating with the ion boundary.

In the case of the popular Tris-glycine system, it is possible to play with the pH in mainly one direction, i.e. lowering it. With a gel pH of 8.8, the system operates at a pH of 9.5, and further increase in the pH results in polyacrylamide hydrolysis, with severe consequences on separation performances. Furthermore, 8.8 is already quite far from the pK of Tris (8.05) so that the buffering power of the system drops at higher pH. Thus the upper limit for use of the glycine-based system and its application for resolving low molecular mass proteins has been reached with the ammediol-glycine system [41]. This system has never reached strong popularity because it operates too close to the basic limits tolerated by acrylamide. Furthermore, it should be kept in mind that basic conditions promote deamidation not only on the polyacrylamide gel, but also at glutamine and asparagine, i.e. conditions that are not favourable for a proteomic analysis.

Conversely, lowering the pH in the tris-glycine system has several advantages such as a better buffering power and a better resolution of high molecular mass proteins, as first exemplified by Johnson [42], and further extended to giant proteins (1MDa) by Fritz et al. [43].

When resolution of low molecular mass proteins at a more neutral pH is needed, it becomes necessary to change the trailing ion, and to opt for chemicals with a lower pK, so that their mobility will be higher for the same pH. The extreme in this direction is represented by the MES system [44] (pK = 6 for MES instead of pK= 9.74 for glycine). However, the most popular systems are based on other Good's buffers, namely Bicine [45] and Tricine [46], as these system show good performances for low molecular mass proteins at a gel pH lesser that 8.5, i.e. with very good buffering power.

In 2D gel systems, the Tris-Tricine system has been successfully used in cases where the resolution of low molecular mass proteins is needed, e.g. for mitochondrial proteins [47], but also for microorganisms [48, 49].

Further developments of the Tricine-based system have been made to replace the glycine-based system. In fact, a Tricine system operating at gel pH of 7 has the same separation window than a glycine system operating at gel pH of 8.8 [50], with the advantages of lesser hydrolysis of amides (and thus longer shelf life for ready to use gels) and of lesser reactivity of nucleophilic amino acids toward spurious chemicals such as residual acrylamide. However, operating at pH 7 also poses buffering problems with Tris, and the situation is further complicated by the fact that many low pK bases, which could replace Tris as gel buffers interfere strongly with polyacrylamide polymerization, either by inhibiting it (e.g. imidazole) or by acting as polymerization catalysts (e.g. triethanolamine), leading to excessive heat production during polymerization and thus irregular gels.

Thus, the two most popular electrophoresis systems operate in their own separate niche, Tricine (pK 8.05) for low molecular mass proteins, and glycine (pK 9.74) for medium and high molecular mass proteins. This suggests that a system using a trailing ion of intermediate pK (i.e. ca. 9) should be able to operate on the whole spectrum with Tris as the only buffer component needed. Indeed, a taurine-based system is able to operate over the whole spectrum, at the expense of a precise control of the gel pH [51] (see figure 2). Here again, 2D electrophoresis of mitochondrial proteins has shown the ability of this system to separate low-molecular mass proteins without increasing acrylamide concentration too much [52]. However, this system is clearly not limited to low molecular mass proteins [53, 54]

An additional, but less understood, phenomenon taking place when changing from one buffer system to another one consists in alterations of the relative mobilities of proteins in the SDS gel. Theoretically, proteins migrate according to their molecular mass in SDS gels. However, there are exceptions to this basic rule in every buffer system, but the ones observed in the widely used glycine system are of course more documented. For example, it has been shown that Tricine and borate-based buffers show an altered migration of some cytoskeletal proteins, and indeed more accurate compared to their molecular mass [50]. This phenomenon can be used to enhance separation of some proteins in some crowded zones in 2D gels, although the results cannot be predicted.

2.3. Variations in the detergent

As the name says, SDS PAGE is carried out in the presence of dodecyl sulfate as the primary solubilizing and charge-conferring agent. The performances of this detergent are documented by more than 40 years of use and thousands of satisfied scientists. However, there are cases in proteomics studies where it would be nice to have even better performances, either in the separation process itself or in the post-separation process, when the SDS gel in interfaced with the process leading to identification by mass spectrometry.

As to the separation itself, there are cases in the course of a proteomics study where the study gets focused on a few proteins, and it is the desirable to separate the proteins of interest from other proteins migrating very close to them, or even to separate variants of the proteins of interest. To this purpose, the detergent used can be altered, on order to induce differential detergent binding and thus differential mobility. This has been observed empirically when comparing different sources of SDS [55], and the concept has been rationalized by mimicking the detergent composition most able to induce separation of normally comigrating proteins [56]. Such alterations of the detergent composition may prove quite efficient in some cases. It must be kept in mind however, that the resolution and performances of SDS electrophoresis rely on a strong binding of numerous detergent molecules to proteins, so that the space for variations is rather limited, as shown in [57].

As to the post-separation steps, SDS, as most detergents, is not the mass spectrometer best friend, and it would be nice use detergents that can be chemically cleaved at will into products that are more mass spectrometry-friendly. Two families of cleavable anionic detergents have been described to date, acid-cleavable detergents [58] and photocleavable ones [59]. Only acid-cleavable detergents have been used to a certain extent. However, their separation properties are much poorer than the ones of SDS [58], as could be anticipated from the stringent constraints applying to detergents for electrophoretic separations [57].

3. Variations in the first dimension: zone electrophoresis

This section of the review will focus on how to improve and simplify zone electrophoretic separations used in conjunction with SDS PAGE to build two-dimensional systems. In order to maximize the resolution, this separation should be as different as possible from the one obtained by SDS PAGE. As the magnitude of this difference is clearly IEF > native zone electrophoresis >> denaturing zone electrophoresis, it is clear that systems using denaturing zone electrophoresis as a first dimension will be used mainly when the other systems do not perform well, e.g. for membrane proteins [60, 61]

As can be seen from [57], electrophoresis in the presence of cationic detergents is clearly the best candidate for denaturing separations, a fact that had been discovered empirically before [3].
Compared to this core protocol, changes and improvements will target the same areas as in SDS PAGE, i.e. gel structure, buffer system and detergent.

Electrophoresis in the presence of cationic detergents proceeds with a reverse polarity compared to SDS gels, and at acidic pH compared to the basic pH used for SDS PAGE. Consequently, polyacrylamide hydrolysis is not a problem in this case. However, gel polymerization cannot proceed with the standard TEMED-persulfate system, as protonated TEMED does not induce persulfate decomposition. In most cases, the Fenton's system described in the original publication [62] is used. However, this system is not as reliable as the TEMED-persulfate one, so that the robust methylene blue-based system [32] can be used very successfully in this case

(see figure 3).

As to the detergent, only two cationic detergents have been successfully used to date, the original benzalkonium chloride, either with a pure C16 chain (16-BAC stands for hexadecyl benzyl dimethyl ammonium chloride) or with a mixture of hydrocarbon chains [62], and CTAB (hexadecyl trimethylammonium bromide) [63]. What has changed over the years, however, is the process for interfacing the first dimension on the second one. In the core protocol [3], the cationic gel was first fixed and stained to remove the cationic detergent, and then equilibrated in SDS prior to loading onto the SDS gel. In more recent protocols, this fixation and staining is omitted, and the gel containing the cationic detergent is just rinsed in water prior to equilibration in SDS buffer [63], or even directly equilibrated in SDS buffer [64].

What remains poorly explored with this technique, however, is the choice of buffers. Because of the presence of the cationic detergent, electrophoresis could theoretically be carried out at any pH. Indeed, in one-dimensional cationic PAGE various buffers have been used, and the very acidic phosphate-glycine buffer used for 16-BAC electrophoresis [62] is not the most widely used. Other systems operating at higher pH, such as the acetate-beta alanine buffer [65] or even the Tricine-arginine buffer [66] have also been described. Indeed, the phosphate-glycine buffer system operates at a pH where the classical calculations for discontinuous systems are no longer valid, so that an electrophoretic system specially designed for electrophoresis in the presence of cationic detergents has been recently proposed [67]. This buffer system uses methoxyacetic as the buffering compound. The separating gel is cast at pH 3 and operates at pH 2.5, and the stacking gel is cast at pH 4 (acetate buffer) and operates at pH 2.9. For comparison, in the phoshate buffer system [62] the separating gel is cast at pH 2 and operates at pH 1.5, and the stacking gel is cast at pH 4 and operates at pH 2.2. However, it has been recently shown that the operative pH seems to play an important role for protein solubility [68], a feature that can be of weak importance when working with soluble proteins, as in [65-67], but this is not necessarily the case with membrane proteins, which are the most important field of application of this technique in proteomics [61].

4. Variations in the first dimension: isoelectric focusing

In a sense, the constraints existing on isoelectric focusing gels are the contrapositive of those existing on SDS PAGE, so that the areas for flexibility and improvement are quite different. In SDS PAGE, the gel is simple to make and many buffers can be used, so that there is a lot of flexibility in these two areas, as shown above. However, there are a lot of constraints on the detergent as the protein-solubilizing agent, so that the flexibility is quite limited in this area.

In isoelectric focusing, the difficult task is to form the pH gradient, and thus the buffers are extremely constrained. Now that immobilized pH gradient dominate the field of isoelectric focusing, these constraints on the buffers have also translated into constraints on the gel, as immobilized pH gradients are by definition grafted into the gel. Just as an example, the standard gel for SDS PAGE is made of just two monomers of quite related structure (acrylamide and Bis). A wide range immobilized pH gradient (e.g. 3-10) uses eight different Immobilines plus the two standard monomers, i.e. ten monomers that must be copolymerized with equal efficiency to ensure the accuracy of the gradient, and this has been shown not to be obvious [69]. Changing the main gel-forming monomer (acrylamide) would mean to reassess the polymerization efficiency of all Immobilines with this new monomer, and it is therefore not surprising that such changes in the monomer have been limited to cases where acrylamide was bound to fail, i.e. isoelectric focusing in very basic pH gradients [70].

Oppositely to the situation of the IEF gels, there is a lot of flexibility in the area of solubilizing agents used for IEF. In fact, on the one hand there is a single chemical constraint, i.e. the solubilizing agents must be non ionic and shall not change the charges of the proteins, and on the other hand it is quite difficult to keep the proteins in solution under such conditions of low ionic strength and with no charge modifications. In fact, isoelectric precipitation is as old as isoelectric focusing (e.g. in [71]). Thus, considerable effort has been devoted to improving this situation, and this has resulted in many additives used to increase protein solubility during isoelectric focusing.

In proteomics, isoelectric focusing is always conducted under denaturing conditions, i.e. inducing protein unfolding. Furthermore, disulfide bridges are also reduced to induce complete dissociation of proteins into separate, unfolded polypeptides.
From the very beginning of 2D gels, protein solubilization has been carried out with neutral chaotropes such as urea [5], and most often used in conjunction with detergents [6]. Thus, the variations that have been brought to this basic protocol have touched the chaotropes and the detergents.

As to chaotropes, there is relatively little choice, chemically speaking, as the chaotrope must remain neutral over the whole pH range. This limits the choice to urea, alkylureas, thiourea, and combinations thereof. Alkylureas proved completely inefficient [72], as could be expected from their weak denaturing power [73]. Oppositely, mixtures of thiourea and urea proved more efficient than urea alone, here again in line with the denaturing power observed in solution [73]. This was shown to be true in every isoelectric focusing setup, i.e. carrier ampholytes IEF in agarose [74], IEF with immobilized pH gradients [72], and carrier ampholytes IEF in polyacrylamide [75]. The latter setup, however, is the most difficult to use, as thiourea inhibits acrylamide polymerization, thereby necessitating the use of the highly efficient methylene blue-based photopolymerization initiator [32, 75].

The situation is quite different for detergents, where any detergent non ionic over the pH range of interest can be used in IEF. Furthermore, the incentive of solubilizing membrane protein in 2D gels has always been a driving force to test non-classical detergents in conjunction with urea as protein solubilizers for 2D gels. As Triton X-100 is known to be one of the best solubilizers of the polyethylene glycol-based detergents [76], the interest has turned to other classes of uncharged detergents, such as glucosides [77], but also members of the sulfobetaine class. The first one to be used, and by far still the most popular, was CHAPS [78]. Linear sulfobetaines, i.e. the most potent nonionic detergents [79], have also been tested. However, as they are not fully compatible with urea, this required decreasing the urea concentration [80]. This poor compatibility (which is temperature-dependent) has considerably impaired their use in 2D gels [81]. To alleviate this problem, special sulfobetaines with improved compatibility were prepared and used [82, 83], at that time mainly with descriptive results. However, when protein identification techniques developed, it became possible to assess more thoroughly the real gain in protein solubilization obtained by various detergents, and especially for membrane proteins. The gains were assessed for sulfobetaines [84, 85], glucosides [86, 87], phosphocholines [88] and even more classical, polyethylene glycol-based detergents [87]. While the positive trends observed earlier on glucosides and sulfobetaines were confirmed by these studies (see an example in figure 4), it was shown quite surprisingly that some of the polyethylene glycol-based detergents perform better than CHAPS in urea plus thiourea chaotrope, while the reverse is true in urea alone.

Despite the documented gains in protein solubility, it is now quite clear that all these variations have proven unable to solubilize a high proportion of the membrane proteins expressed by cells [89].

Besides solubilization of proteins by chaotropes and detergents, the problems caused by cysteines have brought their share of modifications compared to the original protocols. As mentioned earlier, denaturing isoelectric focusing implies reduction of the cysteines in order to separate polypeptidic chains. No problems were encountered as long as carrier ampholytes-based IEF was used, i.e. not extending much higher than pH 7, or when transient focusing systems were used for basic proteins [90]. However, with the use of immobilized pH gradients, it became obvious that cysteine oxidation was a real problem in the basic pH intervals [91]. Indeed, the thiol compounds used for disrupting disulfide bridges and keeping cysteines in their reduced form behave as weak acids, and migrate toward the anode when the pH is high enough. This fact has been used in SDS gels to keep giant proteins reduced [43]. Thus, in an IEF setup, the basic portion of the pH gradient becomes devoid of reducing agents, so that cysteines can reoxidize in many ways, including disulfide bridges, which induces a lot of trailing of the proteins in the basic region of the gels.

To solve this problem, three strategies have been proposed. The first one consists in the continuous infusion of thiol-containing reducers from the cathode [92], so that the proteins are kept in a reducing environment. The obvious problem with this strategy is its absolute need for a careful tuning and timing, in other words its lack of robustness. The second strategy consists in alkylating the cysteines, so that they become chemically inert and cannot either make spurious disulfide bridges nor react on unreacted monomers. Although this strategy seems straightforward and efficient, it has been shown that it does work properly [93]. The issue is that various cysteines on different proteins show quite different reactivities, depending on their environment and thus on the neighbouring amino acids. Consequently, it is almost impossible to find conditions where all the cysteines will be alkylated and none of the other nucleophilic amino acids (e.g. lysine, tyrosine). This unfortunately holds true for a wide range of alkylating agents [94]. Thus, only the third strategy has proven efficient. It consists in blocking the cysteines of the proteins as disulfide bridges, using a vast excess of a low molecular mass disulfide [95]. Because of the very high specificity of the reaction, high concentrations of the disulfide can be used, thereby ensuring maximum blocking of the protein thiols. As an added benefit of this strategy, it is possible to perform a simplified equilibration process, in which no alkylation of the cysteines is needed, opposite to the standard process with immobilized pH gradients [96]. Furthermore, as this process is reversible, either the cysteines can be left protected as disulfides until the mass spectrometry step, using the mass of the adduct as a fixed modification, or the cysteines can be alkylated after spot excision, prior to digestion, extraction and mass spectrometry.

5. Concluding remarks

Prefractionation of samples is generally a pre-requisite in proteomics, and protein fractionation by electrophoresis is quite a versatile and reliable toolbox to perform such prefractionation. As this toolbox has now been used for decades by biochemists and proteomicists, it now relies on a corpus of well-established, robust and efficient techniques and protocols. However, in the course of a proteomic study, there are often cases where it would be very advantageous to be able to tune the separation. This means to increase the resolution and/or the solubilization for proteins of interest, even if it means loosing resolution or solubilization for other proteins that are no longer of interest when the initial proteomics screening process has been achieved.
Numerous ways of tuning electrophoretic separations have been described over the years in the literature, and this review cannot claim to be comprehensive for all these modifications. However, many of them are not widely known nowadays. Thus, the main paths for modifications have been reviewed here, in order to provide the reader the rationale and references to be able to design his own separation system tuned to his own needs, and to adapt to the demands risen by different proteomic projects.

References

[1] Miller I, Eberini I, Gianazza E. Other than IPG-DALT: 2-DE variants. Proteomics.2010; 10:586-610.

[2] Booth AG. Novel System for 2-Dimensional Electrophoresis of Membrane Proteins. Biochemical Journal. 1977;163:165-8.

[3] MacFarlane DE. Two-Dimensional Benzyldimethyl-N-Hexadecylammonium Chloride-]Sodium Dodecyl-Sulfate Preparative Polyacrylamide-Gel Electrophoresis - a High-Capacity High-Resolution Technique for the Purification of Proteins from Complex-Mixtures. Analytical Biochemistry. 1989;176:457-63.

[4] Schägger H, Von Jagow G. Blue Native Electrophoresis for Isolation of Membrane-Protein Complexes in Enzymatically Active Form. Analytical Biochemistry. 1991;199:223-31.

[5] MacGillivray AJ, Rickwood D. The heterogeneity of mouse-chromatin nonhistone proteins as evidenced by two-dimensional polyacrylamide-gel electrophoresis and ion-exchange chromatography. Eur J Biochem. 1974;41:181-90.

[6] O'Farrell PH. High resolution two-dimensional electrophoresis of proteins. J Biol Chem. 1975;250:4007-21.

[7] Wittig I, Schägger H. Features and applications of blue-native and clear-native electrophoresis. Proteomics. 2008;8:3974-90.

[8] Chiari M, Righetti PG. The Immobiline Family - from Vacuum to Plenum Chemistry. Electrophoresis. 1992;13:187-91.

[9] Miller I, Crawford J, Gianazza E. Protein stains for proteomic applications: Which, when, why? Proteomics. 2006;6:5385-408.

[10] Voris BP, Young DA. Very-high-resolution two-dimensional gel electrophoresis of proteins using giant gels. Anal Biochem. 1980;104:478-84.

[11] Patton WF, Lopez MF, Barry P, Skea WM. A Mechanically Strong Matrix for Protein Electrophoresis with Enhanced Silver Staining Properties. Biotechniques. 1992;12:580-5.

[12] Artoni G, Gianazza E, Zanoni M, Gelfi C, Tanzi MC, Barozzi C, et al. Fractionation Techniques in a Hydro-Organic Environment .2. Acryloyl-Morpholine Polymers as a Matrix for Electrophoresis in Hydro-Organic Solvents. Analytical Biochemistry. 1984;137:420-8.

[13] Hochstrasser DF, Patchornik A, Merril CR. Development of Polyacrylamide Gels That Improve the Separation of Proteins and Their Detection by Silver Staining. Analytical Biochemistry. 1988;173:412-23.

[14] Whistler RL, Panzer HP, Roberts HJ. 1-Acrylamido-1-Deoxy-D-Glucitol, 1-Deoxy-1-Methacrylamido-D-Glucitol and Their Polymerization. Journal of Organic Chemistry. 1961;26:1583-8.

[15] Kozulic M, Kozulic B, Mosbach K. Poly-N-Acryloyl Tris Gels as Anticonvection Media for Electrophoresis and Isoelectric-Focusing. Analytical Biochemistry. 1987;163:506-12.

[16] Gelfi C, Debesi P, Alloni A, Righetti PG. Investigation of the Properties of Novel Acrylamido Monomers by Capillary Zone Electrophoresis. Journal of Chromatography. 1992;608:333-41.

[17] Chiari M, Micheletti C, Nesi M, Fazio M, Righetti PG. Towards New Formulations for Polyacrylamide Matrices - N-Acryloylaminoethoxyethanol, a Novel Monomer Combining High Hydrophilicity with Extreme Hydrolytic Stability. Electrophoresis. 1994;15:177-86.

[18] Simò-Alfonso E, Gelfi C, Sebastiano R, Citterio A, Righetti PG. Novel acrylamido monomers with higher hydrophilicity and improved hydrolytic stability .1. Synthetic route and product characterization. Electrophoresis. 1996;17:723-31.
[19] Bendz M, Möller MC, Arrigoni G, Wahlander A, Stella R, Cappadona S, et al. Quantification of membrane proteins using nonspecific protease digestions. J Proteome Res. 2009;8:5666-73.
[20] Donato H, Doig MT, Priest DG. Enhancement of polyacrylamide gel slice dissolution in hydrogen peroxide by cupric sulfate. J Biochem Biophys Methods. 1988;15:331-5.
[21] Gelfi C, Righetti PG. Polymerization Kinetics of Polyacrylamide Gels .1. Effect of Different Cross-Linkers. Electrophoresis. 1981;2:213-9.
[22] Airey JA, Rogers MJ, Sutko JL. Use of a Reversible Polyacrylamide-Gel Cross-Linker in Western Blotting for Rapid Transfer of a Wide Size Range of Polypeptides. Biotechniques. 1991;10:605-8.
[23] Kim YK, Kwon YJ. Isolation of intact proteins from acid-degradable polyacrylamide gel. Proteomics. 2009;9:3765-71.
[24] Righetti PG, Gelfi C, Bosisio AB. Polymerization Kinetics of Polyacrylamide Gels .3. Effect of Catalysts. Electrophoresis. 1981;2:291-5.
[25] Righetti PG, Caglio S. On the kinetics of monomer incorporation into polyacrylamide gels, as investigated by capillary zone electrophoresis. Electrophoresis. 1993;14:573-82.
[26] Bonaventura C, Bonaventura J, Stevens R, Millington D. Acrylamide in polyacrylamide gels can modify proteins during electrophoresis. Anal Biochem. 1994;222:44-8.
[27] Geisthardt D, Kruppa J. Polyacrylamide-Gel Electrophoresis - Reaction of Acrylamide at Alkaline pH with Buffer Components and Proteins. Analytical Biochemistry. 1987;160:184-91.
[28] Hochstrasser DF, Merril CR. 'Catalysts' for polyacrylamide gel polymerization and detection of proteins by silver staining. Appl Theor Electrophor. 1988;1:35-40.
[29] Righetti PG, Chiari M, Casale E, Chiesa C. Oxidation of alkaline immobiline buffers for isoelectric focusing in immobilized pH gradients. Appl Theor Electrophor. 1989;1:115-21.
[30] Brewer JM. Artifact Produced in Disc Electrophoresis by Ammonium Persulfate. Science. 1967;156:256-7.
[31] Chiari M, Micheletti C, Righetti PG, Poli G. Polyacrylamide-Gel Polymerization under Nonoxidizing Conditions, as Monitored by Capillary Zone Electrophoresis. Journal of Chromatography. 1992;598:287-97.
[32] Lyubimova T, Caglio S, Gelfi C, Righetti PG, Rabilloud T. Photopolymerization of Polyacrylamide Gels with Methylene-Blue. Electrophoresis. 1993;14:40-50.
[33] Caglio S, Righetti PG. On the Ph-Dependence of Polymerization Efficiency, as Investigated by Capillary Zone Electrophoresis. Electrophoresis. 1993;14:554-8.
[34] Caglio S, Righetti PG. On the Efficiency of Methylene-Blue Versus Persulfate Catalysis of Polyacrylamide Gels, as Investigated by Capillary Zone Electrophoresis. Electrophoresis. 1993;14:997-1003.
[35] Rabilloud T, Vincon M, Garin J. Micropreparative One-Dimensional and 2-Dimensional Electrophoresis - Improvement with New Photopolymerization Systems. Electrophoresis. 1995;16:1414-22.
[36] Sun G, Anderson VE. Prevention of artifactual protein oxidation generated during sodium dodecyl sulfate-gel electrophoresis. Electrophoresis. 2004;25:959-65.
[37] Lasonder E, Ishihama Y, Andersen JS, Vermunt AM, Pain A, Sauerwein RW, et

al. Analysis of the Plasmodium falciparum proteome by high-accuracy mass spectrometry. Nature. 2002;419:537-42.
[38] Davis BJ. Disc Electrophoresis .2. Method and Application to Human Serum Proteins. Annals of the New York Academy of Sciences. 1964;121:404-27.
[39] Jovin TM. Multiphasic Zone Electrophoresis .1. Steady-State Moving-Boundary Systems Formed by Different Electrolyte Combinations. Biochemistry. 1973;12:871-8.
[40] Jovin TM. Multiphasic Zone Electrophoresis .2. Design of Integrated Discontinuous Buffer Systems for Analytical and Preparative Fractionation. Biochemistry. 1973;12:879-90.
[41] Bury AF. Analysis of Protein and Peptide Mixtures - Evaluation of Three Sodium Dodecyl Sulfate-Polyacrylamide Gel-Electrophoresis Buffer Systems. Journal of Chromatography. 1981;213:491-500.
[42] Johnson BF. Enhanced Resolution in Two-Dimensional Electrophoresis of Low-Molecular-Weight Proteins While Utilizing Enlarged Gels. Analytical Biochemistry. 1982;127:235-46.
[43] Fritz JD, Swartz DR, Greaser ML. Factors Affecting Polyacrylamide-Gel Electrophoresis and Electroblotting of High-Molecular-Weight Myofibrillar Proteins. Analytical Biochemistry. 1989;180:205-10.
[44] Kyte J, Rodriguez H. A Discontinuous Electrophoretic System for Separating Peptides on Polyacrylamide Gels. Analytical Biochemistry. 1983;133:515-22.
[45] Wiltfang J, Arold N, Neuhoff V. A New Multiphasic Buffer System for Sodium Dodecyl Sulfate-Polyacrylamide Gel-Electrophoresis of Proteins and Peptides with Molecular Masses 100 000-1000, and Their Detection with Picomolar Sensitivity. Electrophoresis. 1991;12:352-66.
[46] Schägger H, Von Jagow G. Tricine Sodium Dodecyl-Sulfate Polyacrylamide-Gel Electrophoresis for the Separation of Proteins in the Range from 1-Kda to 100-Kda. Analytical Biochemistry. 1987;166:368-79.
[47] Kruft V, Eubel H, Jänsch L, Werhahn W, Braun HP. Proteomic approach to identify novel mitochondrial proteins in Arabidopsis. Plant Physiology. 2001;127:1694-710.
[48] Hernychova L, Stulik J, Halada P, Macela A, Kroca M, Johansson T, et al. Construction of a Francisella tularensis two-dimensional electrophoresis protein database. Proteomics. 2001;1:508-15.
[49] Fountoulakis M, Juranville JF, Roder D, Evers S, Berndt P, Langen H. Reference map of the low molecular mass proteins of Haemophilus influenzae. Electrophoresis. 1998;19:1819-27.
[50] Patton WF, Chungwelch N, Lopez MF, Cambria RP, Utterback BL, Skea WM. Tris Tricine and Tris Borate Buffer Systems Provide Better Estimates of Human Mesothelial Cell Intermediate Filament Protein Molecular-Weights Than the Standard Tris Glycine System. Analytical Biochemistry. 1991;197:25-33.
[51] Tastet C, Lescuyer P, Diemer H, Luche S, van Dorsselaer A, Rabilloud T. A versatile electrophoresis system for the analysis of high- and low-molecular-weight proteins. Electrophoresis. 2003;24:1787-94.
[52] Lescuyer P, Strub JM, Luche S, Diemer H, Martinez P, Van Dorsselaer A, et al. Progress in the definition of a reference human mitochondrial proteome. Proteomics. 2003;3:157-67.
[53] Linker RA, Brechlin P, Jesse S, Steinacker P, Lee DH, Asif AR, et al. Proteome Profiling in Murine Models of Multiple Sclerosis: Identification of Stage Specific Markers and Culprits for Tissue Damage. Plos One. 2009;4.

[54] Brechlin P, Jahn O, Steinacker P, Cepek L, Kratzin H, Lehnert S, et al. Cerebrospinal fluid-optimized two-dimensional difference gel electrophoresis (2-D DIGE) facilitates the differential diagnosis of Creutzfeldt-Jakob disease. Proteomics. 2008;8:4357-66.
[55] Swaney JB, Vande Woude GF, Bachrach HL. Sodium dodecylsulfate-dependent anomalies in gel electrophoresis: alterations in the banding patterns of foot-and-mouth disease virus polypeptides. Anal Biochem. 1974;58:337-46.
[56] Brown EG. Mixed Anionic Detergent Aliphatic Alcohol Polyacrylamide-Gel Electrophoresis Alters the Separation of Proteins Relative to Conventional Sodium Dodecyl-Sulfate Polyacrylamide-Gel Electrophoresis. Analytical Biochemistry. 1988;174:337-48.
[57] Lopez MF, Patton WF, Utterback BL, Chungwelch N, Barry P, Skea WM, et al. Effect of Various Detergents on Protein Migration in the 2nd-Dimension of 2-Dimensional Gels. Analytical Biochemistry. 1991;199:35-44.
[58] Ross AR, Lee PJ, Smith DL, Langridge JI, Whetton AD, Gaskell SJ. Identification of proteins from two-dimensional polyacrylamide gels using a novel acid-labile surfactant. Proteomics. 2002;2:928-36.
[59] Epstein WW, Jones DS, Bruenger E, Rilling HC. The Synthesis of a Photolabile Detergent and Its Use in the Isolation and Characterization of Protein. Analytical Biochemistry. 1982;119:304-12.
[60] Hartinger J, Stenius K, Hogemann D, Jahn R. 16-BAC/SDS-PAGE: A two-dimensional gel electrophoresis system suitable for the separation of integral membrane proteins. Analytical Biochemistry. 1996;240:126-33.
[61] Braun RJ, Kinkl N, Beer M, Ueffing M. Two-dimensional electrophoresis of membrane proteins. Analytical and Bioanalytical Chemistry. 2007;389:1033-45.
[62] MacFarlane DE. Use of Benzyldimethyl-Normal-Hexadecylammonium Chloride (16-Bac), a Cationic Detergent, in an Acidic Polyacrylamide-Gel Electrophoresis System to Detect Base Labile Protein Methylation in Intact-Cells. Analytical Biochemistry. 1983;132:231-5.
[63] Navarre C, Degand H, Bennett KL, Crawford JS, Mortz E, Boutry M. Subproteomics: Identification of plasma membrane proteins from the yeast Saccharomyces cerevisiae. Proteomics. 2002;2:1706-14.
[64] Helling S, Schmitt E, Joppich C, Schulenborg T, Müllner S, Felske-Müller S, et al. 2-D differential membrane proteome analysis of scarce protein samples. Proteomics. 2006;6:4506-13.
[65] Mòcz G, Balint M. Use of Cationic Detergents for Polyacrylamide-Gel Electrophoresis in Multiphasic Buffer Systems. Analytical Biochemistry. 1984;143:283-92.
[66] Akins RE, Levin PM, Tuan RS. Cetyltrimethylammonium Bromide Discontinuous Gel-Electrophoresis - Mr-Based Separation of Proteins with Retention of Enzymatic-Activity. Analytical Biochemistry. 1992;202:172-8.
[67] Kramer ML. A new multiphasic buffer system for benzyldimethyl-n-hexadecylammonium chloride polyacrylamide gel electrophoresis of proteins providing efficient stacking. Electrophoresis. 2006;27:347-56.
[68] Rabilloud T, Chevallet M, Luche S, Lelong C. Fully denaturing two-dimensional electrophoresis of membrane proteins: A critical update. Proteomics. 2008;8:3965-73.
[69] Righetti PG, Ek K, Bjellqvist B. Polymerization Kinetics of Polyacrylamide Gels Containing Immobilized pH Gradients for Isoelectric-Focusing. Journal of Chromatography. 1984;291:31-42.

[70] Görg A, Obermaier C, Boguth G, Csordas A, Diaz JJ, Madjar JJ. Very alkaline immobilized pH gradients for two-dimensional electrophoresis of ribosomal and nuclear proteins. Electrophoresis. 1997;18:328-37.
[71] Rathnam P, Saxena BB. Isolation and physicochemical characterization of luteinizing hormone from human pituitary glands. J Biol Chem. 1970;245:3725-31.
[72] Rabilloud T, Adessi C, Giraudel A, Lunardi J. Improvement of the solubilization of proteins in two-dimensional electrophoresis with immobilized pH gradients. Electrophoresis. 1997;18:307-16.
[73] Gordon JA, Jencks WP. Relationship of Structure to Effectiveness of Denaturing Agents for Proteins. Biochemistry. 1963;2:47-57.
[74] Oh-Ishi M, Hirabayashi T. Micro-two-dimensional gel electrophoresis with agarose gel in the first dimension. J Chromatogr B Analyt Technol Biomed Life Sci. 1988;32:113-20.
[75] Rabilloud T. Use of thiourea to increase the solubility of membrane proteins in two-dimensional electrophoresis. Electrophoresis. 1998;19:758-60.
[76] Umbreit JN, Strominger.Jl. Relation of Detergent HLB Number to Solubilization and Stabilization of D-Alanine Carboxypeptidase from Bacillus-Subtilis Membranes. Proc Natl Acad Sci U S A. 1973;70:2997-3001.
[77] Witzmann F, Jarnot B, Parker D. Dodecyl Maltoside Detergent Improves Resolution of Hepatic Membrane-Proteins in 2-Dimensional Gels. Electrophoresis. 1991;12:687-8.
[78] Perdew GH, Schaup HW, Selivonchick DP. The Use of a Zwitterionic Detergent in Two-Dimensional Gel-Electrophoresis of Trout Liver-Microsomes. Analytical Biochemistry. 1983;135:453-5.
[79] Navarrete R, Serrano R. Solubilization of Yeast Plasma-Membranes and Mitochondria by Different Types of Non-Denaturing Detergents. Biochimica Et Biophysica Acta. 1983;728:403-8.
[80] Gyenes T, Gyenes E. Effect of Stacking on the Resolving Power of Ultrathin-Layer Two-Dimensional Gel-Electrophoresis. Analytical Biochemistry. 1987;165:155-60.
[81] Satta D, Schapira G, Chafey P, Righetti PG, Wahrmann JP. Solubilization of Plasma-Membranes in Anionic, Non-Ionic and Zwitterionic Surfactants for Iso-Dalt Analysis - a Critical-Evaluation. Journal of Chromatography. 1984;299:57-72.
[82] Gianazza E, Rabilloud T, Quaglia L, Caccia P, Astruatestori S, Osio L, et al. Additives for Immobilized pH Gradient Two-Dimensional Separation of Particulate Material - Comparison between Commercial and New Synthetic Detergents. Analytical Biochemistry. 1987;165:247-57.
[83] Rabilloud T, Gianazza E, Catto N, Righetti PG. Amidosulfobetaines, a Family of Detergents with Improved Solubilization Properties - Application for Isoelectric-Focusing under Denaturing Conditions. Analytical Biochemistry. 1990;185:94-102.
[84] Chevallet M, Santoni V, Poinas A, Rouquié D, Fuchs A, Kieffer S, et al. New zwitterionic detergents improve the analysis of membrane proteins by two-dimensional electrophoresis. Electrophoresis. 1998;19:1901-9.
[85] Rabilloud T, Blisnick T, Heller M, Luche S, Aebersold R, Lunardi J, et al. Analysis of membrane proteins by two-dimensional electrophoresis: Comparison of the proteins extracted from normal or Plasmodium falciparum - infected erythrocyte ghosts. Electrophoresis. 1999;20:3603-10.
[86] Taylor CM, Pfeiffer SE. Enhanced resolution of glycosyl-phosphatidylinositol-anchored and transmembrane proteins from the lipid-rich myelin membrane by two-dimensional gel electrophoresis. Proteomics. 2003;3:1303-12.

[87] Luche S, Santoni V, Rabilloud T. Evaluation of nonionic and zwitterionic detergents as membrane protein solubilizers in two-dimensional electrophoresis. Proteomics. 2003;3:249-53.
[88] Babu GJ, Wheeler D, Alzate O, Periasamy M. Solubilization of membrane proteins for two-dimensional gel electrophoresis: identification of sarcoplasmic reticulum membrane proteins. Analytical Biochemistry. 2004;325:121-5.
[89] Rabilloud T. Membrane proteins and proteomics: love is possible, but so difficult. Electrophoresis. 2009;30 Suppl 1:S174-80.
[90] O'Farrell PZ, Goodman HM, O'Farrell PH. High-Resolution 2-Dimensional Electrophoresis of Basic as Well as Acidic Proteins. Cell. 1977;12:1133-41.
[91] Altland K, Becher P, Rossmann U, Bjellqvist B. Isoelectric-Focusing of Basic-Proteins - the Problem of Oxidation of Cysteines. Electrophoresis. 1988;9:474-85.
[92] Hoving S, Gerrits B, Voshol H, Muller D, Roberts RC, van Oostrum J. Preparative two-dimensional gel electrophoresis at alkaline pH using narrow range immobilized pH gradients. Proteomics. 2002;2:127-34.
[93] Galvani M, Hamdan M, Herbert B, Righetti PG. Alkylation kinetics of proteins in preparation for two-dimensional maps: A matrix assisted laser desorption/ionization-mass spectrometry investigation. Electrophoresis. 2001;22:2058-65.
[94] Luche S, Diemer H, Tastet C, Chevallet M, Van Dorsselaer A, Leize-Wagner E, et al. About thiol derivatization and resolution of basic proteins in two-dimensional electrophoresis. Proteomics. 2004;4:551-61.
[95] Olsson I, Larsson K, Palmgren R, Bjellqvist B. Organic disulfides as a means to generate streak-free two-dimensional maps with narrow range basic immobilized pH gradient strips as first Product dimension. Proteomics. 2002;2:1630-2.
[96] Görg A, Postel W, Weser J, Gunther S, Strahler JR, Hanash SM, et al. Elimination of Point Streaking on Silver Stained Two-Dimensional Gels by Addition of Iodoacetamide to the Equilibration Buffer. Electrophoresis. 1987;8:122-4.

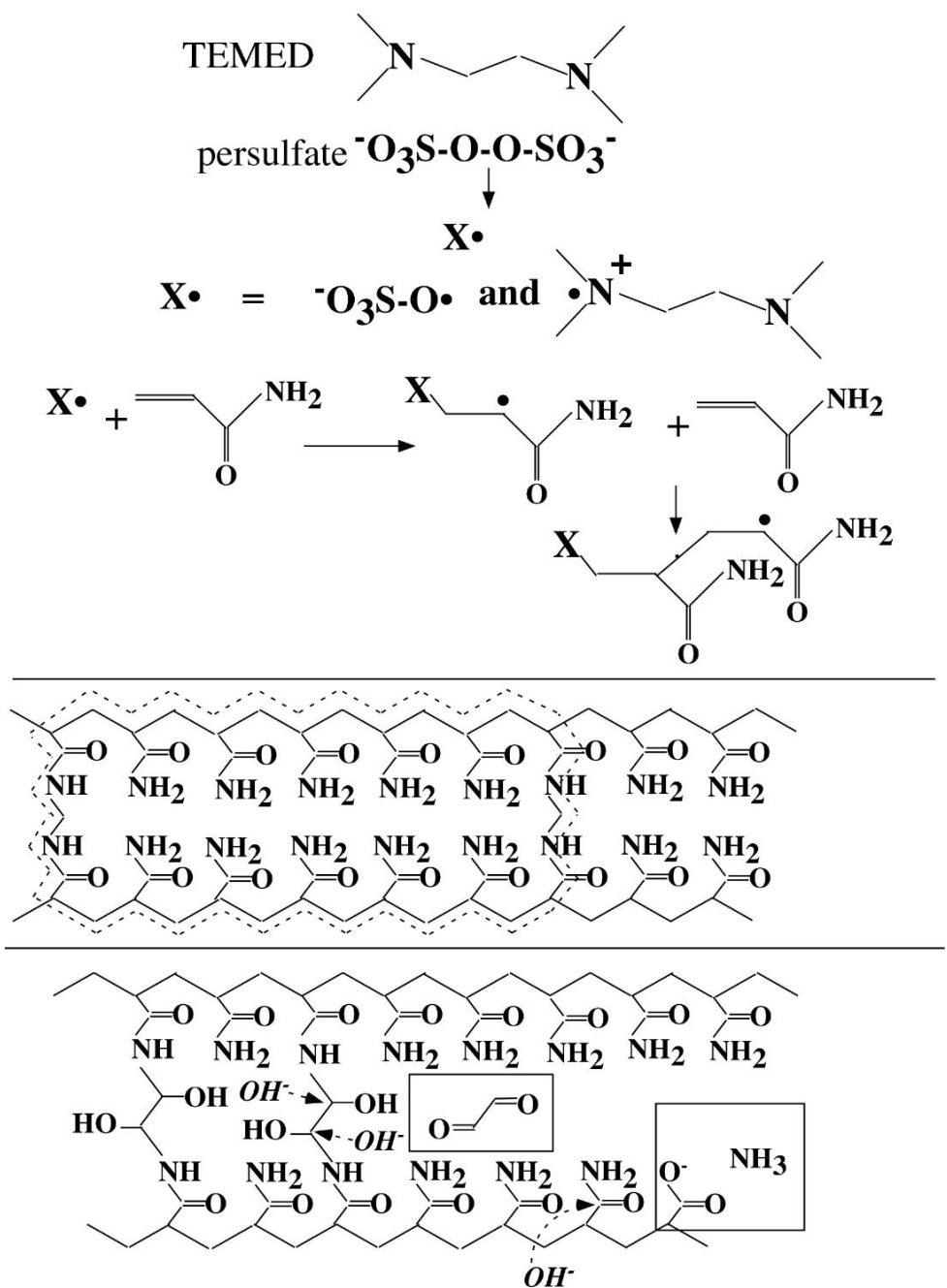

Figure 1: scheme of a polyacrylamide gel

The figure is divided in three panels. In the top panel, the chemistry of polymerization is shown, with the radical production by the TEMED-persulfate couple, and the radical polymerization of acrylamide

In the medium panel, the structure of the gel is shown (Bis crosslinker). The polyethylene backbone and the numerous pending amido groups are visible on this scheme, and a gel pore is circled in dashed line

In the bottom panel, base-induced degradation phenomena are shown (DHEBA-crosslinker). The chemical species induced by the hydrolysis (glyoxal, pending carboxylates) are boxed

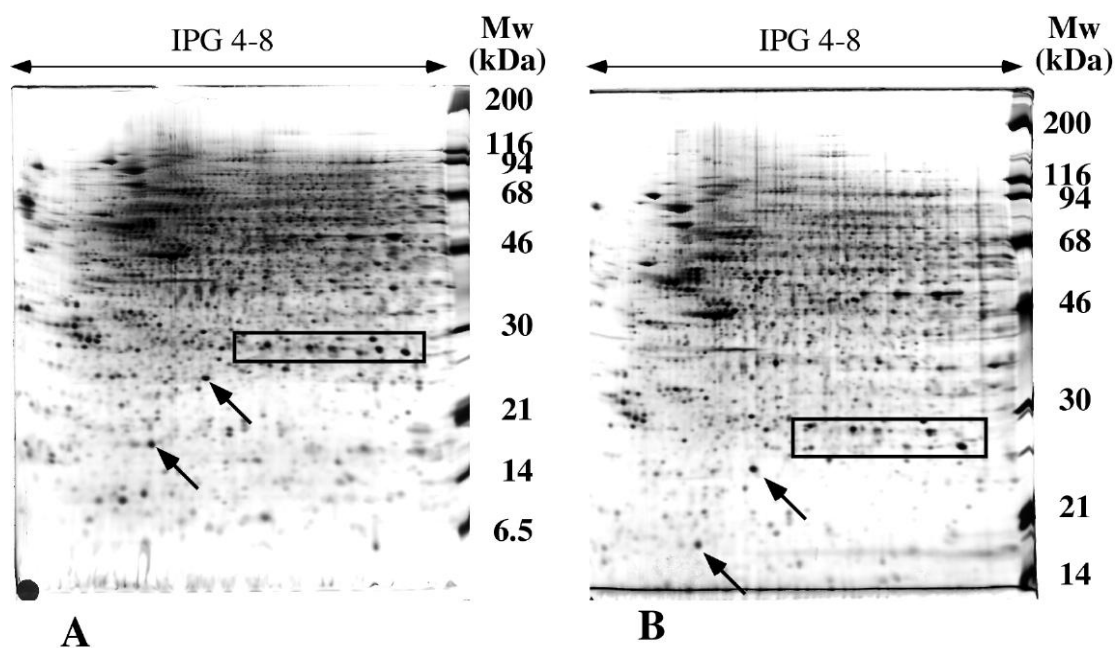

Figure 2: Separation modulation in the SDS dimension by a pH change
120 µg of proteins extracted from HeLa cells were used as a sample, and separated on a linear 4-8 immobilized pH gradient (70,000 Vh). The strips were then loaded on top of a 10% polyacrylamide gel and electrophoresed in a Tris-HCl-Taurine system. Detection by silver staining. Molecular mass markers were loaded on the side of the SDS gel.
Panel A : gel buffer: Tris-HCl pH 8.05
Panel B: gel buffer: Tris-HCl pH 7.75
A few equivalent proteins on both gels are shown by a box and arrows

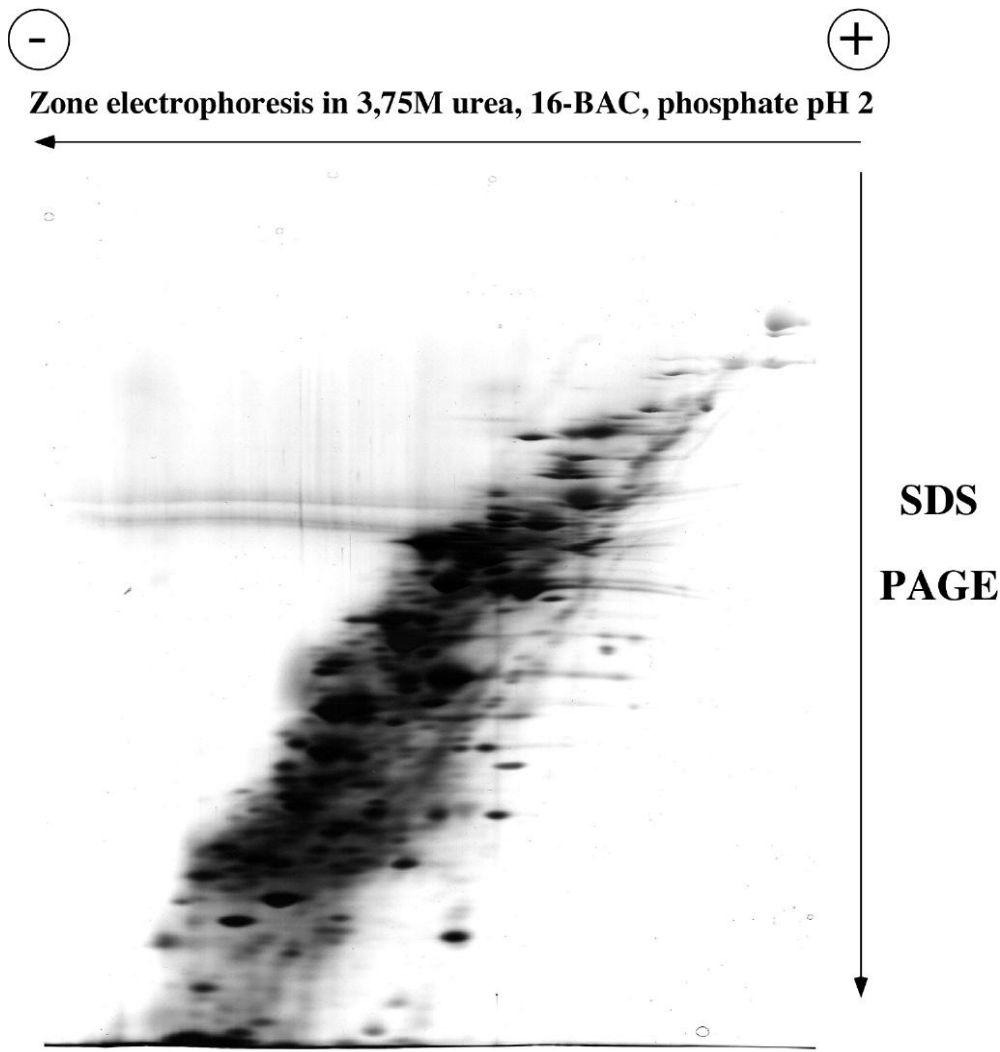

Figure 3: separation of membrane proteins by double zone electrophoresis
A membrane preparation from B. subtilis was used as the sample. 100µg proteins, solubilized in 3.75M urea, 2% 16-BAC, 5mM tris-carboxyethylphosphine, 100mM phosphate buffer pH3, were loaded on the first dimension rod gel, containing 7.5 % acrylamide, 3.75M urea, 0.1% 16-BAC, 200mM phosphate buffer pH 2.1, and photopolymerized with 1mM sodium toluene sulfinate, 20 µM diphenyliodonium chloride and 30µM Methylene Blue. The electrode buffer contained phosphate, glycine and 0.1% 16-BAC

Pyronin Y was used as a tracking dye for the end of the first migration (polarity indicated on the figure). The rod gel was then extruded and equilibrated in high SDS buffer (2.5%) for 15 minutes, before sealing on top of the second dimension, SDS PAGE gel (10% acrylamide). Detection by silver staining. The limited streaking and numerous spots show the efficiency of the whole method, including the polymerization process.
16-BAC stands for hexadecyl benzyl dimethyl ammonium chloride.

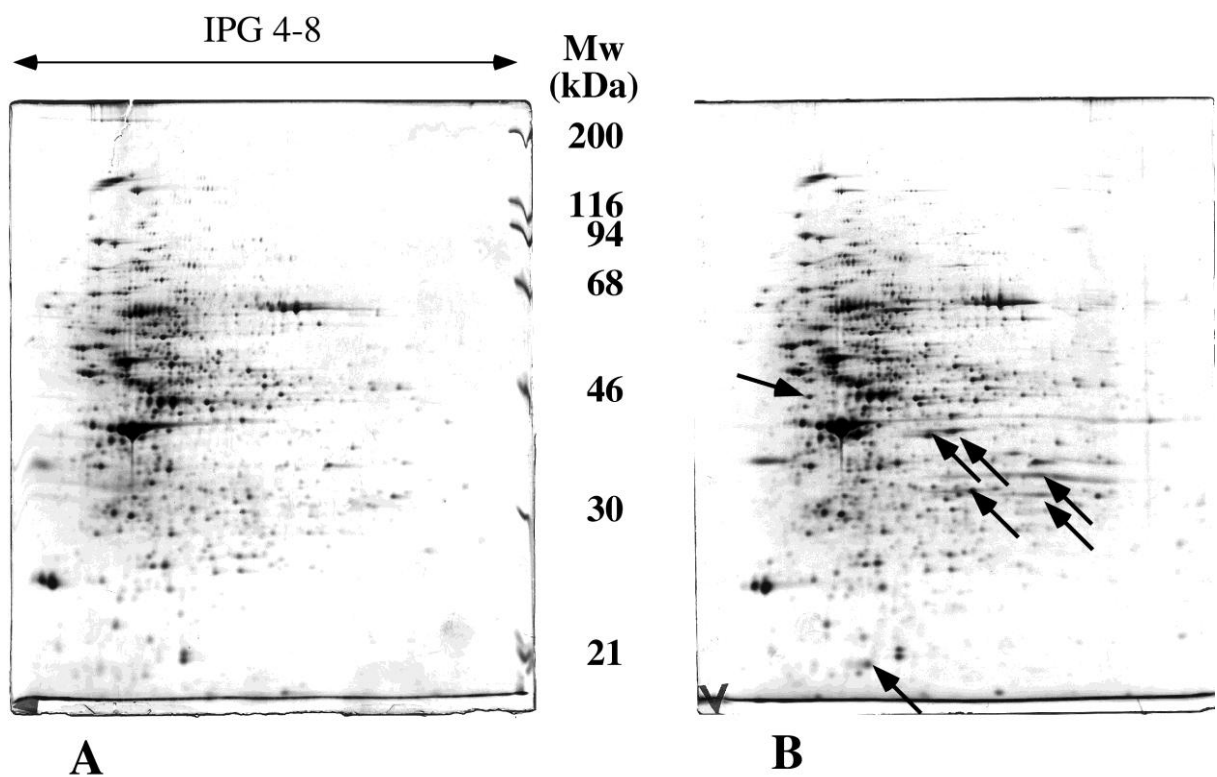

Figure 4: differential protein extraction by detergents in IEF-based 2D electrophoresis
A membrane preparation from B. subtilis was used as the sample. 100µg proteins, solubilized in 7M urea, 2M thiourea, 2% detergent, 0.4% carrier ampholytes and 50mM DTT, were separated by IEF in immobilized pH gradients (same buffer as for extraction, linear 4-8 pH gradient). Migration for 70,000 Vh
Second dimension: 10% acrylamide gel.
Panel A: extraction and migration using CHAPS as the detergent
Panel B: extraction and migration using Brij56 as the detergent
Spots solubilized and focused by Brij56 and not by CHAPS are shown by arrows